\documentclass[journal]{IEEEtran}
\usepackage{amsmath,amssymb,amsfonts,amsthm,enumerate}
\usepackage{hyperref}
\usepackage{subfigure}
\usepackage{cleveref}
\usepackage{graphicx}
\usepackage{epstopdf}
\usepackage{colortbl}

\theoremstyle{plain}

\theoremstyle{definition}

\theoremstyle{remark}

\newcommand{\by}{{\textbf y}}

\begin{document}
\title{Universal Anomaly Detection: Algorithms and Applications}
\author{Shachar~Siboni and~Asaf~Cohen,~\IEEEmembership{Member,~IEEE}
\thanks{Parts of this work appeared at the Workshop on Information Forensics and Security, WIFS 2014, Atlanta, GA. \newline S. Siboni and A. Cohen are with the Department of Communication System Engineering, Ben-Gurion University, Beer-Sheva, 84105, Israel. E-mails: sibonish@bgu.ac.il; coasaf@bgu.ac.il.  \newline Partially supported by the Israeli Chief Scientist under the Kabarnit consortium.}}
\maketitle
\thispagestyle{empty}
\begin{abstract}
Modern computer threats are far more complicated than those seen in the past. They are constantly evolving, altering their appearance, perpetually changing disguise. Under such circumstances, detecting known threats, a fortiori zero-day attacks, requires new tools, which are able to capture the essence of their behavior, rather than some fixed signatures.

In this work, we propose novel universal anomaly detection algorithms, which are able to learn the normal behavior of systems and alert for abnormalities, without any prior knowledge on the system model, nor any knowledge on the characteristics of the attack.   
The suggested method utilizes the Lempel-Ziv universal compression algorithm in order to optimally give probability assignments for normal behavior (during learning), then estimate the likelihood of new data (during operation) and classify it accordingly. 

The suggested technique is generic, and can be applied to different scenarios. Indeed, we apply it to key problems in computer security.  
The first is detecting Botnets Command and Control (C\&C) channels. A Botnet is a logical network of compromised machines which are remotely controlled by an attacker using a C\&C infrastructure, in order to perform malicious activities. We derive a detection algorithm based on timing data, which can be collected without deep inspection, from open as well as encrypted flows. We evaluate the algorithm on real-world network traces, showing how a universal, low complexity C\&C identification system can be built, with high detection rates and low false-alarm probabilities. Further applications include malicious tools detection via system calls monitoring and data leakage identification. 
\end{abstract}
\begin{IEEEkeywords}
Computer Security; Anomaly Detection; Universal Compression; Probability Assignment; Individual Sequences; Botnets; Command and Control Channels; Malicious Tools; Data Leakage.
\end{IEEEkeywords}
\section{Introduction}
\IEEEPARstart{C}{yber-attacks} are a disturbing security threat existing today in communication- and computer-based systems. They affect a wide range of domains including electricity and water infrastructures, financial and capital markets, medicine and healthcare, army, businesses, enterprises and universities around the world. The majority of massive cyber-attacks today are conducted by Botnets, including Distributed Denial-of-Service (DDoS) attacks, spamming, fraud and identity theft, etc. 

A Botnet is a logical network of compromised machines, Bots, which are remotely controlled by a Botmaster using a Command and Control (C\&C) infrastructure. The compromised machines can be any collection of vulnerable hosts, e.g. computers, mobile-phones or tablets. Infection is via infected websites, file-sharing networks, email attachments, and more (see infection tree analysis in \cite{wang2012characteristics}). Once a host is infected and becomes a Bot, it is programmed to use a C\&C channel for further downloads and updates and awaits instructions from the Botmaster. It updates its data and operates upon receiving commands from the Botmaster (e.g., launch a DDoS attack).

The C\&C channel plays a key role in a Botnet by operating as the communication means within the network. The Botmaster manages and controls its Bots using these C\&C channels in order to perform malicious activities on selected targets. This way, the Bots act as a distributed attack platform on-demand, coordinated by the Botmaster. However, due to the fact that the C\&C channels are the only way the Botmaster can communicate with its Bots, they can be considered as \emph{the weakest link of a Botnet, as blocking them, renders the Botnets useless}. Accordingly, a main objective is to identify and block C\&C activities before any real harm is caused.

In order to mask their activities and bypass defense mechanisms such as firewalls, Botnets uses common communication protocols as their C\&C, including IRC \cite{strayer2008botnet,gu2008botsniffer}, HTTP \cite{gu2008botsniffer}, Peer-to-Peer (P2P) \cite{chang2009p2p,noh2009detecting,francois2011botcloud} and DNS \cite{villamarin2008identifying}. Recently, Botnets also adopted social networks as the underlying C\&C \cite{burghouwt2011towards}. However, while it is tempting to develop protocol-specific methods to detect Botnets, attackers constantly improve their C\&C infrastructures and develop new evasion capabilities, including changing signatures of the C\&C traffic, employing encryption and obfuscation and using domain generation \cite{shin2012large} in order to deceive detection systems.

Current techniques for Botnet study and detection are based on honeynets, signatures-based detection and anomaly detection models \cite{silva2013botnets}. Honeynets act as traps in order to collect information about Bots and study their behavior \cite{abbasi2012classification}. Once the mechanism of the monitored Bots is exposed, it is possible to design a designated detection and blocking mechanism. Signature-based approaches rely on a signature database of notorious Botnets that were previously learned. However, signature-based techniques are \emph{prone to zero-day attacks and require a constant update of the signatures} database \cite{silva2013botnets}. Anomaly-based detection techniques, on the other hand, aim to detect anomalies in network traffic or system behaviour, which may indicate the presence of malicious activities. 

A basic assumption when using anomaly detection is that attacks differ from normal behavior. Thus, traffic analysis is used on both packet and flow levels, considering metrics such as rate, volume, latency, response time and timestamps in order to identify anomalous data. Indeed, anomaly detection seems as a promising approach for Botnet detection since it may detect new structures of attacks (zero-day attacks). However, this may come at the cost of high false-alarm rates. Moreover, to achieve good performance, one may require prior knowledge, e.g., statistical assumptions on the normal data, such as a Markov Model \cite{noh2009detecting} or ARMA modeling \cite{celenk2010predictive}.
   
Gu \textit{et al.} proposed two anomaly-based detection systems, BotSniffer \cite{gu2008botsniffer} and BotMiner \cite{gu2008botminer}, based on traffic analysis. The former was design for IRC- and HTTP-based Botnets, while the latter was designed as protocol independent detection, which requires no prior knowledge. However, both systems rely on Deep Packet Inspection techniques, hence are less suitable for on the fly analysis of large amounts of traffic.

AsSadhan \textit{et al.} \cite{assadhan2009detecting} suggested that a periodic behavior indicates Botnet activity. Tegeler \textit{et al.} \cite{tegeler2012botfinder} presented a detection system, BotFinder, considering high-level statistical features of C\&C communication which were extracted from known Botnets in a controlled environment, limiting the ability to detect new types of attacks. Protocol specific systems were given by Villamarn-Salomn and Brustoloni \cite{villamarin2008identifying} for DNS traffic, Chang and Daniels \cite{chang2009p2p} for P2P Botnets topology and Strayer \textit{et al.} \cite{strayer2008botnet} using correlation of packet sizes and timing patterns for IRC-based Botnets.

From classification-based point of view, Lu \textit{et al.} \cite{lu2009botcop} presented a classification scheme, BotCop, using a decision tree statistical model to classify different types of applications. Yet, BotCop applied payload signatures techniques. Este \textit{et al.} \cite{este2009support} and Mazzariello and Sansone \cite{mazzariello2009anomaly} employed a Support Vector Machine as a single-class classifier model which constructs a statistical model based on a given training set in order to distinguish between normal and malicious activities.

The suggested system does not rely on memoryless features of the data, such as specific values or signatures. In contrast, it builds a \emph{context tree} for the learned data, hence, when a new data sequence is tested, the \emph{order} of values or events in it has the main impact on the classification performance.

\subsection{Our Contribution}
We study the problem of detecting Botnets C\&C channels. We suggest a novel universal anomaly detection algorithm, which uses no a-priori information about neither the Botnets traffic patterns nor the normal behavior patterns, yet efficiently learns the normal behavior in order to generate a statistical model to which tested traffic can be compared.  

Our classification model is based on the celebrated Lempel Ziv algorithm, which is known as an optimal universal compression algorithm and hence a preferable universal prediction algorithm when applied to stationary and ergodic sources over finite alphabets. Using the probability assignment induced by the prediction algorithm, we rigorously define the statistical model which represents the normal behavior, and offer a mechanism to test new, unknown sequences, using this model. Furthermore, we offer a new look on the \emph{way to use data in the classification process}, offering the \emph{context} of the data sequence as the key characteristic used in the classification. 

We evaluate the model with \emph{real-world} network traces, when timing data is the main tested feature. This allows us to be both protocol-independent and encryption-independent. Moreover, it allows us to suggest a system which is immune to various hiding techniques, especially when used with low-level features of the data, such as timings or sizes. The results clearly show that the suggested model is a favorable solution for the problem at hand, with excellent results in terms of low false alarm and high detection rates, yet with only moderate (linear time) complexity and no deep packet inspection.

Finally, we note that the suggested scheme is applicable to \emph{any sequence of behaviors, and not necessarily only timing data}. Hence, one can use it to test for anomalous application behavior, anomalous communication patterns within an organization and outside it, etc.

A short conference version of this work appeared in \cite{7084311}. The current paper includes algorithms and results for two new applications, together with additional explanations and discussions.

The rest of the paper is organized as follows. \Cref{sec. prelim} gives the required background material. \Cref{sec. sys} describes the key concept of universal anomaly detection, with the key application, \emph{Botnet Identification}, as the main example. \Cref{sec. res} gives the tests results for this case, using real network traces. \Cref{sec. more apps} gives two additional applications, together with results on real data. \Cref{sec. discussion} discusses the possible prevention strategies attackers can use against the suggested system, and proves its robustness by arguing such strategies would require huge amounts of data and massive learning. Finally, \Cref{sec. conc} concludes this paper.  

\section{Preliminaries}\label{sec. prelim}
Classification refers to the problem of labeling unknown (new) instances to the most appropriate class among a set of (known) predefined classes. When the underlying probability distributions  for the classes $\{p_i\}_{i=1}^{M}$ are known, and we wish to decide which generated a given sequence $\by$, a decision rule of the form $\hat{i} = \text{argmax}_{1 \leq i \leq M} p_{i}(\by)$
is optimal in the sense of minimizing the probability of error.
In unary-class classification, however, information is available only on one type of instances. The goal may be to either identify such instances, or, in the case of \emph{anomaly detection}, identify instances which \emph{do not fit the ones learned from}. Indeed, when only few, if any at all, anomalous instances exist to learn from, yet instances of normal behavior are available, one can build a behavioral model \emph{based on the normal instances} and classify any instance deviating from that model as anomalous \cite{chandola2009anomaly}.

Thus, given the probability distribution of the normal data, $p(\cdot)$, the optimal decision rule in terms of maximizing the detection probability given a fixed false alarm probability (in the Neyman-Pearson sense) is to compare $p(\by)$ to a \emph{threshold}, and decide that $\by$ is normal if $p(\by)$ is above the threshold and anomalous otherwise. The threshold is determined according to the required false alarm probability. 
In practice, the underlying distribution which generates the normal sequences is, of course, unknown. A reasonable approach in this case is to estimate it using the previously observed sequences and use the resulting estimate $\hat{p}(\cdot)$. Note, however, that the estimation problem differs significantly if a statistical model (e.g., i.i.d.\ or Markovian of a certain order) is given, if the only knowledge is that the sequences are related to some stationary and ergodic source, or, in the ``worst" case, the data constitutes of \emph{individual sequences}, that is, deterministic sequences with no pre-defined statistical model.

In this paper, we suggest an anomaly detection technique for the most general case, where no underlying statistical model is given. To do this, we build on the relation between \emph{prediction} of discrete sequences and \emph{lossless compression} \cite{feder1992universal}, in order to use \emph{universal compression algorithms} and their associated \emph{probability assignment} in the anomaly detection procedure.  
\subsection{Universal Probability Assignment}\label{prob assign}
The Lempel Ziv algorithm \cite{ziv1978compression}, LZ78, is a universal compression algorithm with a vanishing \emph{redundancy}. Consequently, it can also be used as an optimal \emph{universal prediction} algorithm \cite{feder1992universal}, using the appropriate probability assignment. The LZ78 algorithm is widely used in a variety of other applications. In the context of classification, it was also used in \cite{nisenson2003towards} for typist identification based on keyboard events and in \cite{begleiter2004prediction} for English text, music pieces and proteins classification.
For completeness, we briefly describe the compression method and the associated probability assignment algorithm. 

The LZ78 algorithm is a dictionary-based compression method. For a given sequence of data symbols, a dictionary of phrases parsed from that sequence is constructed based on the incremental parsing process as follows. At the beginning the dictionary is empty. Then, during each step of the algorithm, the smallest prefix of consecutive data symbols not yet seen, i.e., which does not exist in the dictionary, is parsed and added to the dictionary. By that, each phrase is a unique phrase in the dictionary, that may extend a previously seen phrase by one symbol.

Given a sequence $s_{1}^n = (s_{1}s_{2}\ldots s_{n})$, a \emph{parsed phrase}, $P$, is the smallest prefix of consecutive data symbols that has not been seen yet. This can also be considered as suffix concatenation of symbol $s_{i}$ (from the sequence) with a previously seen phrase $P'$ (from the dictionary), i.e., $P = (P's_{i})$. A \emph{dictionary}, $D$, is a collection of all distinct phrases parsed from a given data sequence$s_1^n$, i.e., $D = \{P_{1}, P_{2}, \ldots, P_{i}, \ldots P_{n} \}$. For example, the sequence $aabdbbacbbda$ is parsed as $a|ab|d|b|ba|c|bb|da|$.

A common representation of the dictionary is a rooted-tree where each phrase in the dictionary is represented as a path from the root to an internal node in the tree according to the set of symbols the phrase consists of. In addition, leaf-nodes are added as suffix for each phrase in the tree. A statistical model can be defined for a given data sequences during the construction of a phrase-tree  \cite{feder1992universal}, as described next. 

At the beginning, an initial tree is constructed including only a root node and $k$ leaf-nodes as its children, where $k$ is the size of the alphabet. Then, for each new phrase parsed from a sequence, the tree is traversed, starting from the root, following the set of symbols the phrase consists of, and ending at the appropriate leaf-node. Once a leaf-node is reached, the tree is extended at this point by adding all the symbols from the alphabet as immediate children nodes to that leaf, making it an internal node.  In order to define a statistical model, each node in the tree, except for the root node, maintains a node traversal counter, where each leaf-node's counter is set to 1 and each internal node's counter is equal to the sum of its immediate children's counters. 

\begin{figure*}[tb]
\centering
\includegraphics[width=\textwidth]{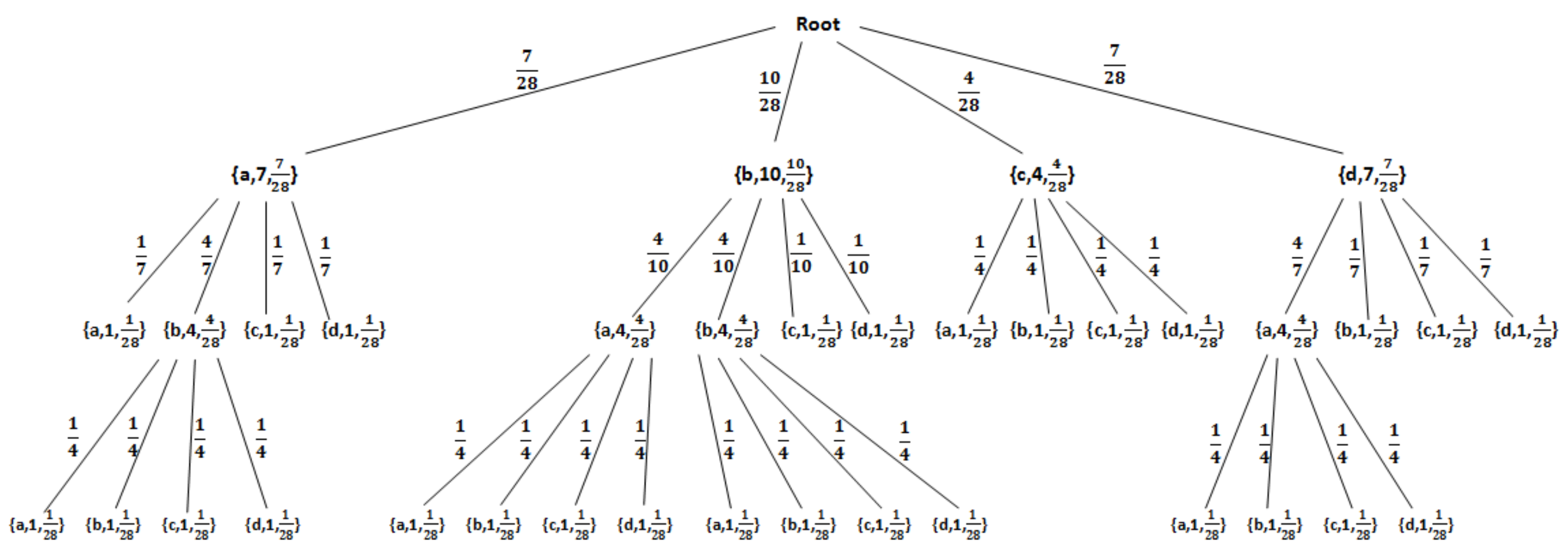}
\caption{An LZ78 Statistical Model for sequence ``aabdbbacbbda''.}
\label{fig:lz78_stat_model}
\end{figure*}

For a probability assignment, as all leaf-nodes' counters are set to $1$, they are assumed uniformly distributed with a probability  $1/i$, where $i$ is the total number of leaf-nodes. Each internal node's probability is defined as the sum of its immediate children's probabilities, which also equals the ratio between its counter and current $i$. For example, Figure~\ref{fig:lz78_stat_model} demonstrates the resulting statistical model for the sequence ``aabdbbacbbda". Each node in the tree is represented by the 3-tuple $\{$symbol, counter, probability$\}$. In addition, the probability of an edge is defined by dividing the nodes' probabilities. Note that the probabilities of edges connected directly to the root are equal to the appropriate root-children's counter divided by the total number of leaf-nodes, $i$, at each step of the algorithm. 
The probability of a phrase $P_{i} \in D$ is calculated by multiplying the probabilities of the edges along the path defined by the symbols of $P_{i}$. Moreover, note that for each phrase $P_{i}$ there exist a specific node in the tree whose probability represents the probability of that phrase. For instance, from the example shown in Figure~\ref{fig:lz78_stat_model}, it can be seen that $P(ba) = \frac{10}{28} \times \frac{4}{10} = \frac{4}{28}$. Considering a sequence $S$, if during the traversal a leaf-node is reached before all the symbols of $S$ are finished then the traversal return to the root and continue until all the symbols of that sequence are consumed \cite{nisenson2003towards}. For example, the probability of the sequence ``bdca" given the same statistical model above, is defined as the following traversal probabilities multiplication: Root$\rightarrow$b$\rightarrow$d$\rightarrow$Root$\rightarrow$c$\rightarrow$a and is calculated as:
\[
P(bdca|M_{aabdbbacbbda}) = \frac{10}{28} \times \frac{1}{10} \times \frac{4}{28} \times \frac{1}{4} = \frac{1}{784}.
\]
This stems from the conditional probability $\hat{P}(s_{t+1}|s_{1}^t)$, where $s_{t+1}$ is the next symbol after the (sub-)sequence $s_{1}^t$, which is calculated as the ratio between the counter of symbol $s_{t+1}$ and the counter of symbol $s_{t}$. We consider $s_{1}^t$ as \emph{the context} of $s_{t+1}$ at time $t+1$. 

\section{Anomaly Detection Via Universal Probability Assignment}\label{sec. sys}
We now describe the building blocks of the anomaly detection system. Throughout, the system is described in the context of \emph{detecting anomalies in network traffic}. Thus, in our problem domain, the data instances are discrete sequences of network traces. However, as previously mentioned, the proposed system is generic, and can easily be adapted to detect anomalies in any discrete sequence of events, with the proper preprocessing. In fact, two additional applications of the algorithms below are given in \Cref{sec. more apps}.  
\subsection{Preprocessing}
A data sequence is defined as a series of events from a flow between a specific client and a specific host. The $i$th \emph{Network Event}, denoted by $e_{i,xy}$, is a data transaction between client $x$ and host $y$ and is defined by the tuple $e_{i,xy} = (t_{i}, tt_{i}, csb_{i}, scb_{i},x, y)$,
where $t_{i}$ is the time event $e_{i,xy}$ occurred; $tt_{i}$ is the duration of event $e_{i,xy}$; $csb_{i}$ and $scb_{i}$ are the total number of bytes which were sent by client $x$ to host $y$ and by host $y$ to client $x$, respectively. A \emph{Network Flow}, denoted by $f_{xy}$, is series of network events between client $x$ and host $y$ sorted by their time of occurrence, $t_i$. That is, $f_{xy} = \{e_{1,xy}, e_{2,xy}, \ldots,  e_{n,xy}\}$.

For actual learning and testing, it is not required to use all features (fields) in the data. As shown in the experimental results, good detection capabilities can be achieved even when focusing on a single feature. For example, timing data can be characterized by the \emph{difference between two consecutive events} of the same flow, denoted by Time-Difference (TD) and defined by $TD_{i,xy}=e_{i+1,xy}(t_{i+1})-e_{i,xy}(t_{i})$. A different perspective is the total time the event took, denoted by Time-Taken (TT) and defined by $TT_{i,xy}=e_{i,xy}(tt_{i})$.
Similarly, one can focus only on sizes, e.g., Client-Server-Bytes (CSB) and Server-Client-Bytes (SCB) and respectively define $CSB_{i,xy}=e_{i,xy}(csb_{i})$ and $SCB_{i,xy}=e_{i,xy}(scb_{i})$. 

Consequently, a single-feature data sequence is a serialization of one of the above features, e.g, with respect to Time-Difference, a sequence/flow is defined as: 
\begin{multline*}
f_{xy,TD} = \{e_{2,xy}(t_{2})-e_{1,xy}(t_{1}),\ e_{3,xy}(t_{3})-e_{2,xy}(t_{2}),\\
 \ldots, e_{n,xy}(t_{n})-e_{n-1,xy}(t_{n-1})\}.
\end{multline*}

The above procedure may result in a sequence over a very large alphabet (as, for example, times are given with a very high precision). To reduce the range of values, quantization is performed. For $k$ quantization levels, a set of $k$ centroids $\{c_{1}, c_{2}, c_{3}, \ldots, c_{k}\}$, is used. The centroids are extracted from the available data during the training phase. Clearly, the number of centroids and the method for extracting them may affect the overall results. However, as seen in the experiments on real data, this fine-tuning is easily done during training.

\subsection{Learning}
\begin{figure}
\centering
\includegraphics[width=3.5in]{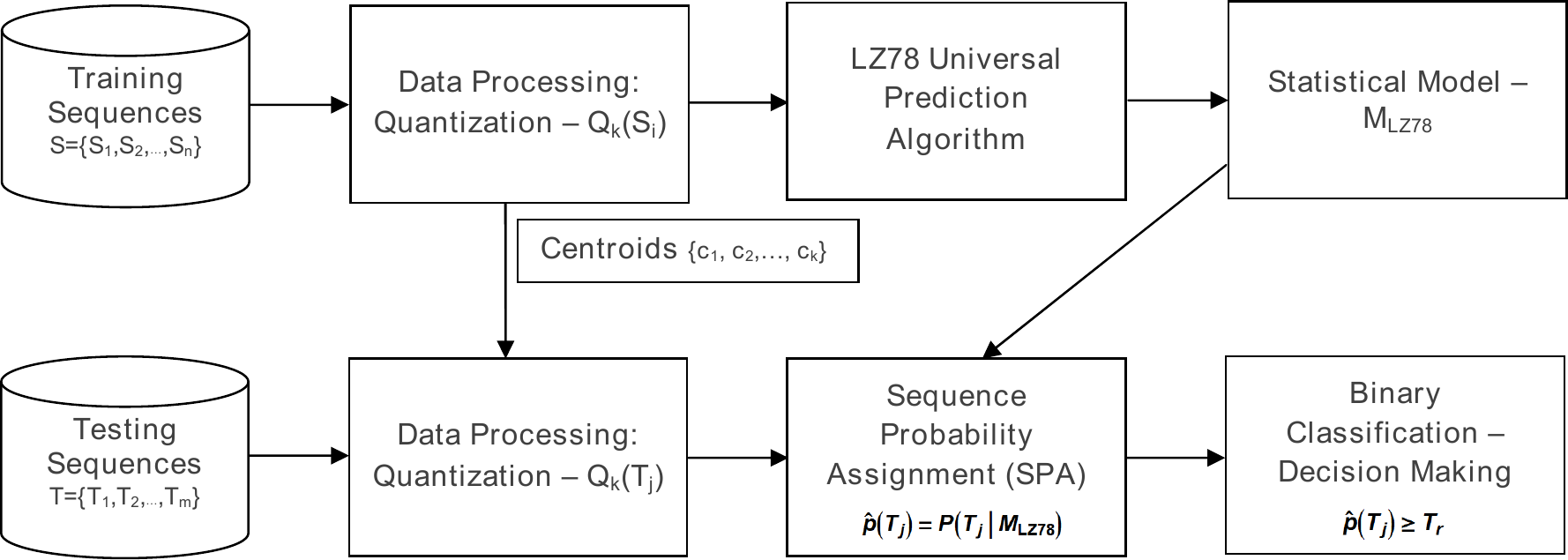}
\caption{A Classification Model based on the LZ78 prediction algorithm}
\label{fig:class_model_lz78}
\end{figure}

The LZ78-based classification model is divided into a learning phase and a testing phase, as illustrated in Figure~\ref{fig:class_model_lz78}. In the learning phase, an LZ78 statistical model is built based on a given training set of discrete (quantized) sequences over finite alphabet $S = \{S_{1}, S_{2}, \ldots, S_{n}\}$, using the mechanism explained in \Cref{prob assign}. Of course, training is done only on normal, benign traffic. 
In the testing phase, first, each testing sequence is separately quantized using \emph{the same quantization method and the same set of centroids} $\{c_{1}, c_{2},\ldots, c_{k}\}$ which were extracted in the learning phase. Then, the probability of each suspected sequence (testing sequence) $T_{j}$ from a given testing set $T = \{T_{1}, T_{2},\ldots, T_{m}\}$ is estimated based on the constructed statistical model (during the learning phase) and classified respective to a predefined threshold $T_{r}$. Specifically, the probability of each testing sequence, $p(T_{j})$, is estimated using sequential probability assignment \emph{given the LZ78 statistical model built in the learning phase}. Testing sequences for which $\hat{p}(T_{j})$ is greater than or equal to $T_{r}$ are classified as normal (as they ``fit" the model) while a lower than threshold value is classified as anomalous. 

Accordingly, the performance of the classifier is measured by the false alarm and hit detection ratios, also known as false positive rate (FPR) or Type 1 error and true positive rate (TPR), respectively, and demonstrated by a ROC (Receiver Operating Characteristic) curve. The false alarm ratio reflects the number of negative instances incorrectly classified as positive in proportion to the total number of negatives in the test, whereas hit detection ratio measures the proportion between the number of positive instances correctly classified and the total number of positives in the test. The ROC curve is generated with respect to a set of thresholds, in our case in the range $[\min_j(\hat{p}(T_{j})), \max_j(\hat{p}(T_{j}))]$, where each threshold defines a specific confusion matrix and results in a specific point in the graph.

Note that the above classification model can be updated, either by extending the existing statistical model or rebuilding it, with new (training) data sequences at any given time. This can be applied once the overall performance get lower than a given threshold for a configurable interval of time (and not for each momentary decrease).

\section{Botnets Identification}\label{sec. res}
The data set used contains high-level real-world network traces provided by an Internet security company (in order to maintain the confidentiality of the company's customers, the name of the company is withheld). The data set consists of 3,714,238 client-server transactions, taken during a time window of approximately 3 hours in specific day in 2009. The whole data set is available at https://dl.dropboxusercontent.com/u/13592090/Botnet2009.rar. Each client, denoted by `cid', may connect with several hosts (web-servers), denoted by `hid'. On each transaction, data is sent both by the client and the host. This defines communication pairs CID\_HID. Each transaction is labeled as either legal, denoted by `good', for normal data traffic generated by the client, or illegal transaction, denoted by `hostile', that is Bots. Labeling was done by the security company's experts based on well-known black-lists. Note that these labels are not used during the classification process. They are used only in the validation phase. Table~\ref{tab:db_struct} exemplify the structure of the data set. Note that each transaction is represented by a single record in the data set, which consists of the following fields: `time', referring to the time the transaction took place; `time-taken', is the total time the transaction took; `cs-bytes' and `sc-bytes' fields represent the total bytes sent by the client/server(host) to the server(host)/client during the transaction, respectively; `mime-type' denotes the Internet content type of the transaction, such as: plain text, image, html page, application, etc.; `cat' is the category of the transaction - `good' or `hostile'; and the `hid' and `cid' fields refer to the host-index (Internet site, web-server) and client-index respectively. Again, to protect the identity of the company's customers, these indices were given arbitrary. However, some malicious sites are identified by their domain name, e.g., `hotsearchworld.com' or `blitzkrieg88.bl.funpic.de'. 
\begin{table}
\centering
\caption{Example for the Database Structure. Each record in the data set represent a data transaction between specific client and specific host/server.}
\scalebox{0.65}{
	\begin{tabular}{ |l|l|l|l|l|l|l|l| }
	\hline
	time & time-taken & cs-bytes & sc-bytes & mime-type & cat & hid & cid 	\\ \hline
	\rowcolor[gray]{0.9} 05:52:37 & 40 & 803 & 360 & image/gif & good & 49 & 9 				\\ \hline
	05:52:37 & 74 & 734 & 277 & text/html & good & 102 & 15 			\\ \hline
	\rowcolor[gray]{0.7} 05:52:37 & 27 & 578 & 507 & image/gif & good & 52 & 486 			\\ \hline
	05:52:37 & 27 & 578 & 507 & image/gif & good & 75526 & 486			\\ \hline
	05:52:37 & 25 & 655 & 4196 & image/jpeg & good & 52 & 4			\\ \hline
	05:52:37 & 25 & 655 & 4196 & image/jpeg & good & 75526 & 4			\\ \hline
	\rowcolor[gray]{0.7} 05:52:37 & 26 & 577 & 505 & image/gif & good & 52 & 486			\\ \hline
	05:52:37 & 26 & 577 & 505 & image/gif & good & 75526 & 486			\\ \hline
	05:52:37 & 31 & 624 & 960 & image/gif & good & 52 & 6				\\ \hline
	05:52:37 & 31 & 624 & 960 & image/gif & good & 75526 & 6			\\ \hline
	05:52:37 & 1 & 812 & 22672 & application/octet-stream & good & 52 & 6	\\ \hline
	05:52:37 & 1 & 812 & 22672 & application/octet-stream & good & 75526 & 6	\\ \hline
	05:52:37 & 30 & 707 & 4368 & image/jpeg & good & 52 & 2			\\ \hline
	05:52:37 & 30 & 707 & 4368 & image/jpeg & good & 75526 & 2			\\ \hline
	05:52:37 & 28 & 667 & 2639 & image/jpeg & good & 52 & 4			\\ \hline
	05:52:37 & 28 & 667 & 2639 & image/jpeg & good & 75526 & 4			\\ \hline
	05:52:37 & 180 & 434 & 1451 & text/html;\%20charset=iso-8859-1 & hostile & 3 & 6 \\ \hline
	05:52:37 & 34 & 710 & 4270 & image/jpeg & good & 52 & 2			\\ \hline
	05:52:37 & 34 & 710 & 4270 & image/jpeg & good & 75526 & 2			\\ \hline
	\rowcolor[gray]{0.9} 05:52:37 & 69 & 697 & 334 & text/css & good & 49 & 9				\\ \hline
	\end{tabular}	
}
\label{tab:db_struct}
\end{table}

Processing of the data included serialization and feature extraction: first, the given data set is split into set of flows based on CID\_HID connections, as illustrated in Table~\ref{tab:db_struct} with respect to flows 486\_52 and 9\_49 (marked in gray). A `Flow' is a sequence of related transactions of the same communication pair CID\_HID sorted by time and with the same label, either `good' or `hostile'. In total, there are 19164 flows labeled as `good' and only 65 `hostile' flows (0.338$\%$). This indicates the imbalance of the data set where most of the transactions are legal and only a small fraction is illegal. However, this is characteristic of real network traffic behavior.

Next, selected features are extracted from each transaction, e.g., Time-Difference, Time-Taken, Server-Client-Bytes and Client-Server-Bytes. After quantization, the resulting sequences are the discrete time, finite alphabet sequence on which learning and testing was performed.

Note that flows from a single client may consist of both `good' flows reflecting legitimate data traffic generated by the client itself, as well as `hostile' flows which are generated by a Bot installed on the client. In contrast, a server (host) has only flows with the same label, that is, if a host was labeled as a C\&C infrastructure than all its transactions are considered malicious. The `Client' and `Host' definitions represented two different perspectives of the data set. On the one hand one can examine the events occurring in the network from the client point of view, and on the other from the host point of view, as will be shown in the following results. 
\subsection{Experiments}
Several experiments were conducted using the above datasets. Flows/sequences were randomly selected from the datasets, from both `Client' and `Host' perspectives, and \emph{divided equally between the training and the testing phases} (hence, ROCs are based only on newly seen data). 

We first tested which single-feature achieves the best results. The system was then optimized using this feature alone. The best results, in terms of optimal threshold and ROC-AUC (Area Under Curve), were achieved using the Time-Difference (TD) representation of the data sequences along with the `Uniform' quantization (several quantization algorithms were tested). To better understand why TD was superior, consider a \emph{legitimate web surfer compared to a hostile connection using HTTP only as a C\&C channel}. While the surfer must have a reasonable behavior in the time domain, affected by the times required to read a page, the times required for the server to respond, etc., a C\&C channel may behave differently, without, for example, a reasonable response time from the server as it only collects data from the bots, and the ``GET" messages are used solely to \emph{transmit} information. Due to space limitation, we do not include the results for the inferior features, and focus on the results under TD and uniform quantization.   

\begin{figure*}[t!]
\centering 
\begin{subfigure}
\centering 
\includegraphics[height = 2.17in]{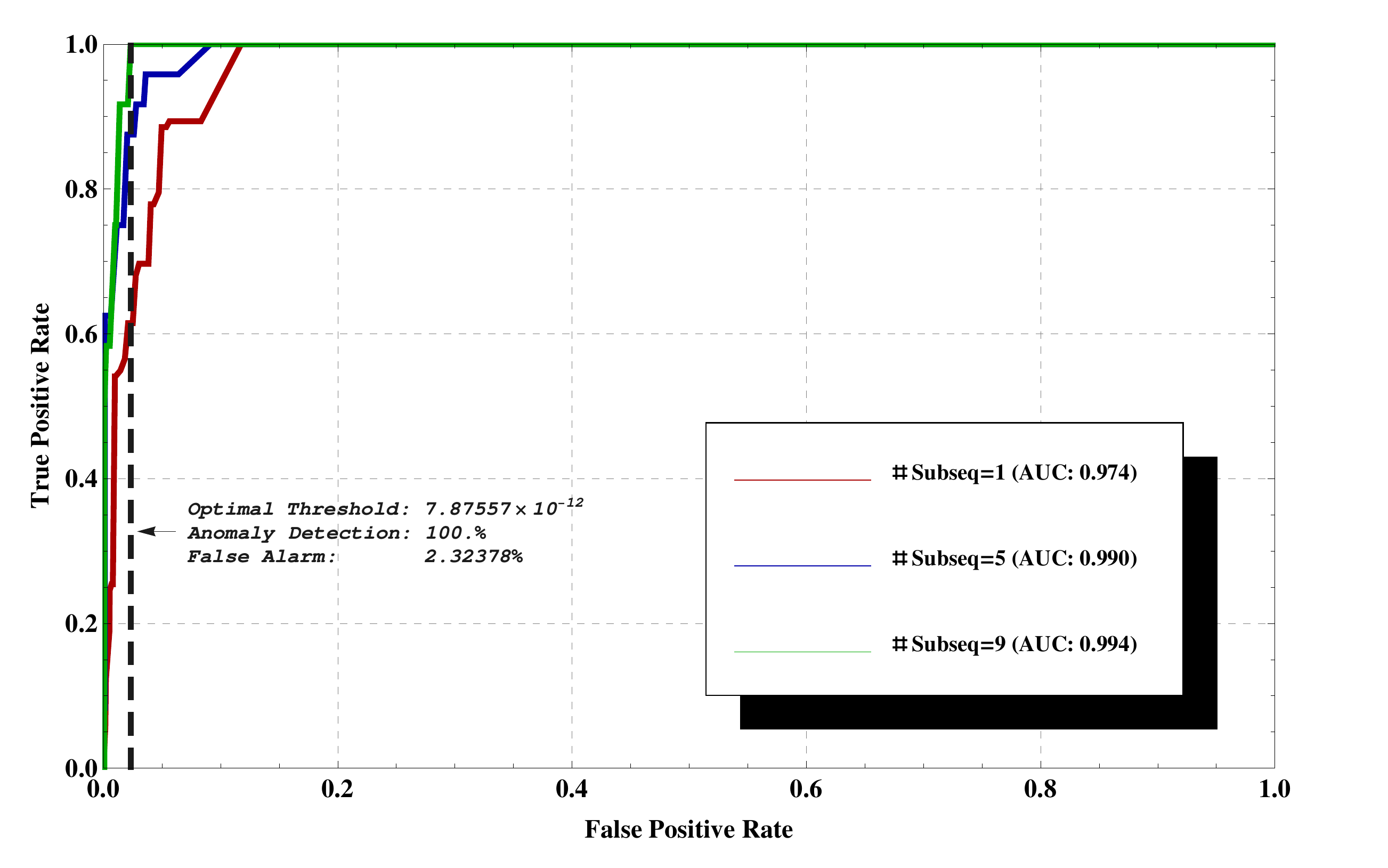}
\end{subfigure}
~
\begin{subfigure}
\centering 
\includegraphics[height = 2.17in]{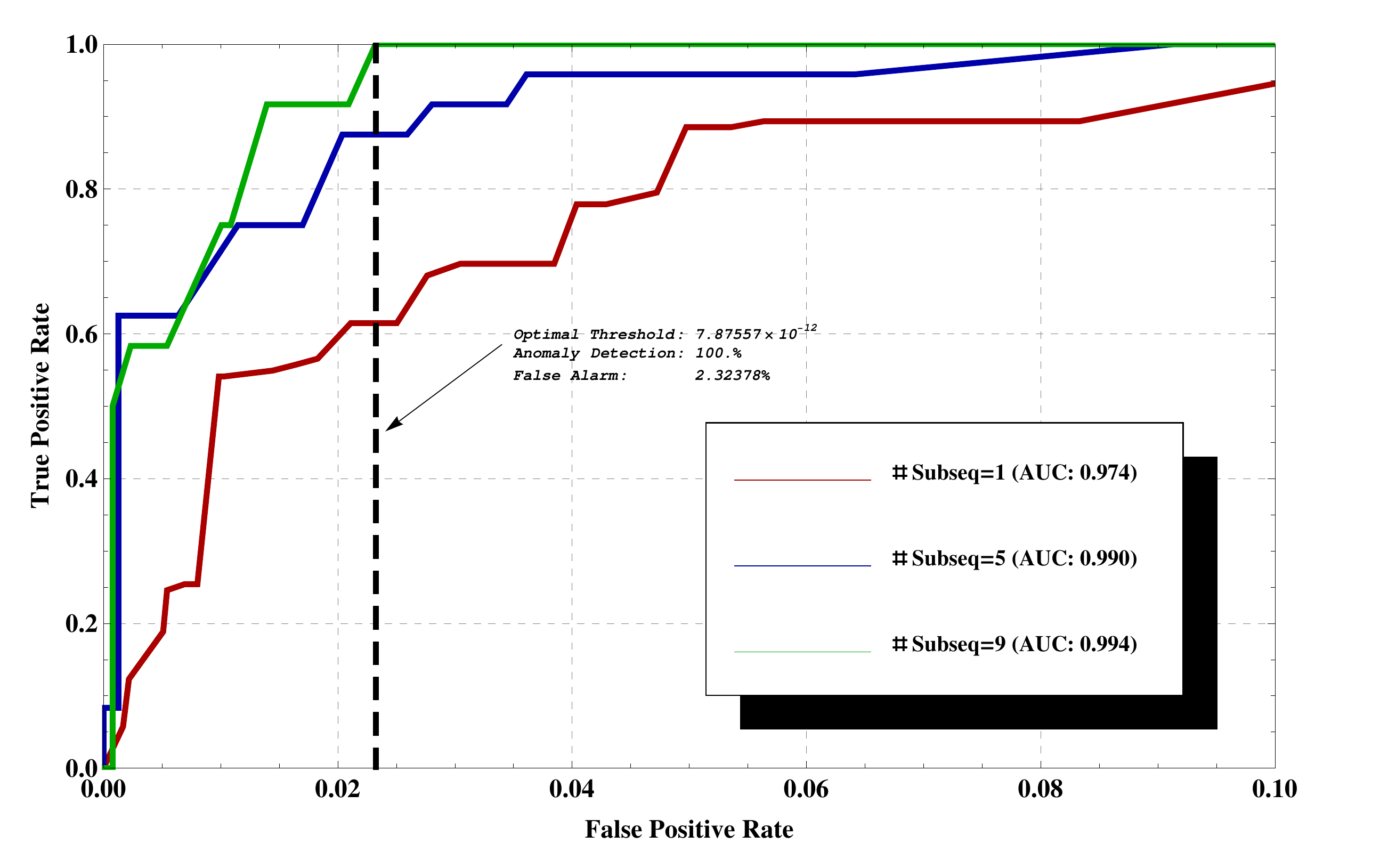}
\end{subfigure}
\caption{Testing `Majority Vote Classification' using TD feature and `Uniform' quantization method by considering `Clients' type of flows. Left: Receiver Operating Curve; Right: Zoom in on the upper left corner. First, each testing sequence partitioned into several sets of subsequences, denoted as $\#Subseq$ in the graph, and the decision made per set of subsequences, where better results achieved for higher subsequences definition.}
\label{fig:cid_lz78_results_Majority}
\end{figure*}

Still, using TD, the optimal threshold for 100\% detection results in 11.75\% false alarms. However, \emph{this is when only a single, short sequence is tested}. To further improve the above results, a majority vote for several sequences within the flow can be used. Each data segment is partitioned into several subsequences of length 10. The classification is done based on the majority of these subsequences' estimations, as either positive or negative, resulting in better classification performance as the number of subsequences is higher. For example, an AUC of 0.994 and false alarms rate of 2.32378\% are achieved using a threshold of 7.87557x10$^{-12}$, as illustrated in Figure~\ref{fig:cid_lz78_results_Majority}. This is obtained at the cost of higher detection time per data segment, of course, as using only one subsequence per data segment the decision is made immediately, while using 9 subsequences causes delay. 

\begin{figure}
\centering 
\includegraphics[scale=0.36]{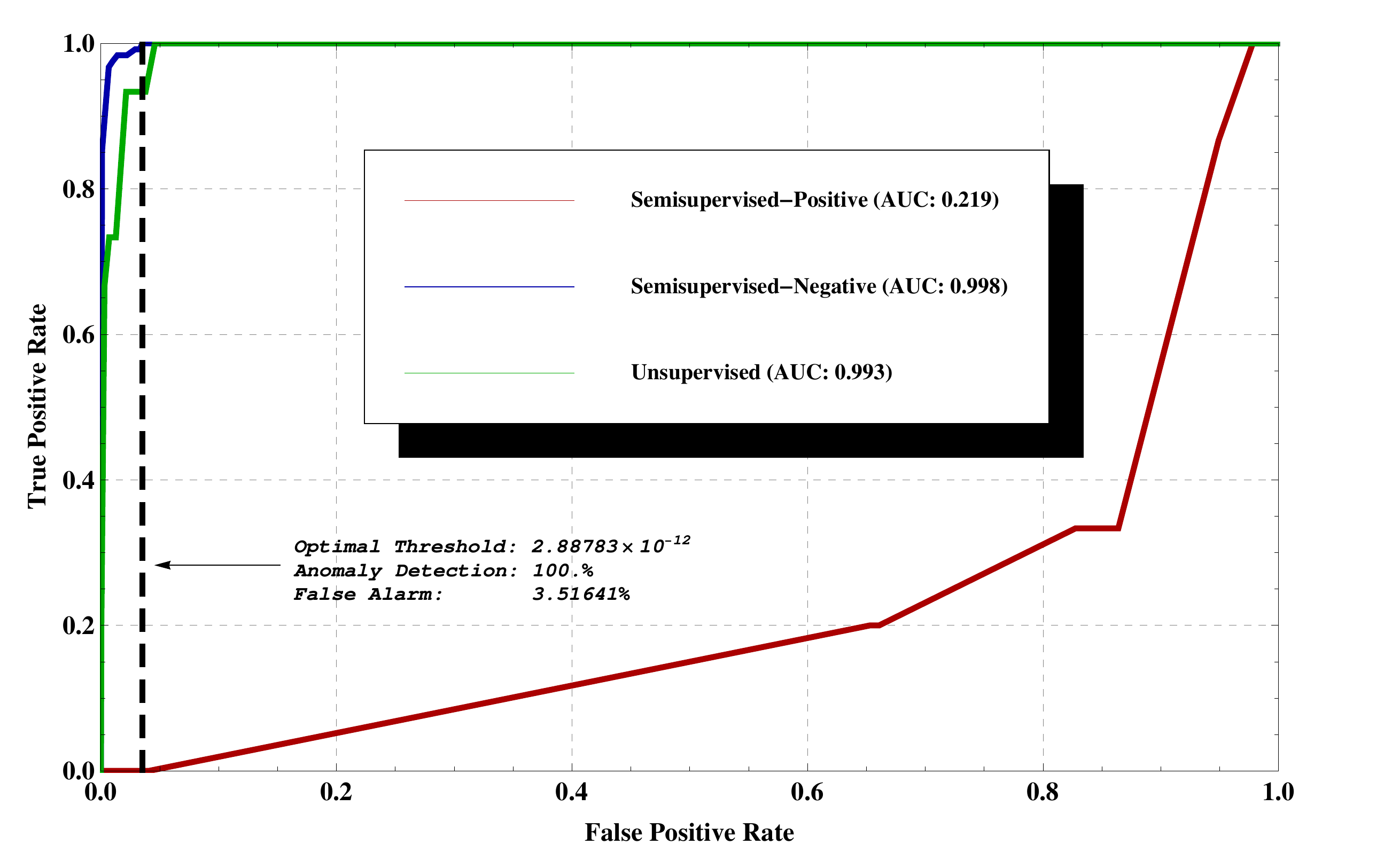}
\caption{Testing `Training Modes': Semisupervised-Negative, Semisupervised-Positive and Unsupervised modes, using TD feature and `Uniform' quantization method with respect to `Hosts' type of flows. The classifier achieves the best results of AUC=0.998 with 100\% detection and 3.51641\% false alarms for Semisupervised-Negative training mode and the worst results of AUC=0.219 and $\sim$98\% false alarms for 100\% detection in case of Semisupervised-Positive training mode.}
\label{fig:cid_lz78_results_HID_LearnTypes}
\end{figure}

The above results were obtained from a `Client' point of view by considering a semisupervised training. To examine the `Host' point of view, and differentiate between `Semisupervised-Negative', where one considers \emph{only normal data sequences during training}, `Semisupervised-Positive', taking into account \emph{only anomalies} sequences during training, and `Unsupervised' where the training set consists of both normal \emph{and few anomalous} sequences as well, we refer the reader to Figure~\ref{fig:cid_lz78_results_HID_LearnTypes}. Clearly, trying to learn the anomalous behavior fails (the red curve - AUC=0.219), as there are only few samples, and C\&C traffic may differ significantly for new bots on which the system was not trained. The key message to take from the figure is, however, that \emph{when learning is done using noisy data, which includes some C\&C traffic besides the normal one, there is no significant degradation in performance.} 
That is, in `Unsupervised' training mode, the classifier achieves very good results despite the fact that the underlying datasets used in the training phase contains both normal and few anomalous sequences. `Semisupervised-Negative' achieves the best results of AUC=0.998. Note that `Unsupervised' is the more realistic scenario where no a priori information is available on the training data.

Finally, for a concrete example, examining flows 6\_3, 6\_14 and 9\_1 under TD, `Uniform' quantization, and a threshold of 2.88783x10$^{-12}$ (obtained from the last test case), we found that all these flows are classified as anomalies. From clients point of view, this indicates that clients 6 and 9 are infected by a Bot program, and form a host perspective this implies that hosts 1, 3 and 14 act as command and control servers. By examine these three servers with a list of the actual domains corresponding to the host our findings were confirmed. Server 1 is known as `hotsearchworld.com', Server 3 is `blitzkrieg88.bl.funpic.de' and Server 14 has IP address of 209.123.8.198, which was black-listed.

\section{Detection of Malicious Tools and Data Leakage}\label{sec. more apps}
In this section, we include additional results which further strengthen the applicability of the universal anomaly detection suggested. Specifically, we apply the anomaly detection system suggested to system calls in order to detect malicious tools on a Windows machine, and to TCP traffic of a server in order to detect unwanted data leakage. In both experiments, the capability of the tool to detect abnormal behaviour without prior knowledge is demonstrated.
\subsection{Monitoring the Context of System Calls for Anomalous Behaviour}
The sequence of systems calls used by a process can serve as an identifier for the process behaviour and use of resources. Moreover, when a program is exploited or malicious tools are running, the sequence of system calls may differ significantly compared to normal behaviour, incriminating the program or entire machine (see, e.g., \cite{fava2008projecting} and references therein). 

In this part of the work, the universal anomaly detection tool was used to learn the \emph{context of normal system calls}, and alert for anomalous behaviour. Specifically, the sequences of system calls created by a process (e.g., firefox.exe) were recorder, processed, and learned. Then, when viewing new data from the same process, the anomaly detection algorithm compared the processed new data to the learned model in order to decide whether the process is still benign, or was it maliciously exploited by some tool. 

Due to the large amount of possible system calls, calls were grouped into 7 types, based on the nature of the call: \emph{Device, Files, Memory, Process, Registry, Security and Synchronization}. That is, unlike the time-difference data described in \Cref{sec. res}, herein, the quantization process did not include any minimization of distances or a requirement for uniform probabilities, but, rather, labeled the calls based on their known functionality. Recording and classification used NtTrace \cite{nttrace}.

In the learning phase, system calls were recorder, quantized according to the types above and then a discrete sequence over the alphabet of size 7 was created. The sequence was used to build the (normal behaviour) LZ tree, as described in \Cref{sec. prelim}, from which a histogram for the probabilities of tuples of length 20 was calculated. This histogram was \emph{the only data saved from the learning phase}. The learning phase included 4 hours of data. 

For testing, segments of 2 minutes were recorded. For each segment, a histogram was calculated, similar to the learning phase (calculating probabilities for tuples of length 20 oven an alphabet of size 7). In this part of the work, decisions were made based on the Kullback-Leibler divergence (the \emph{KL distance} \cite[Section 2.3]{C10}) between the normal histogram and the tested one.  

\begin{figure*}[tb]
\centering 
\includegraphics[width=\textwidth]{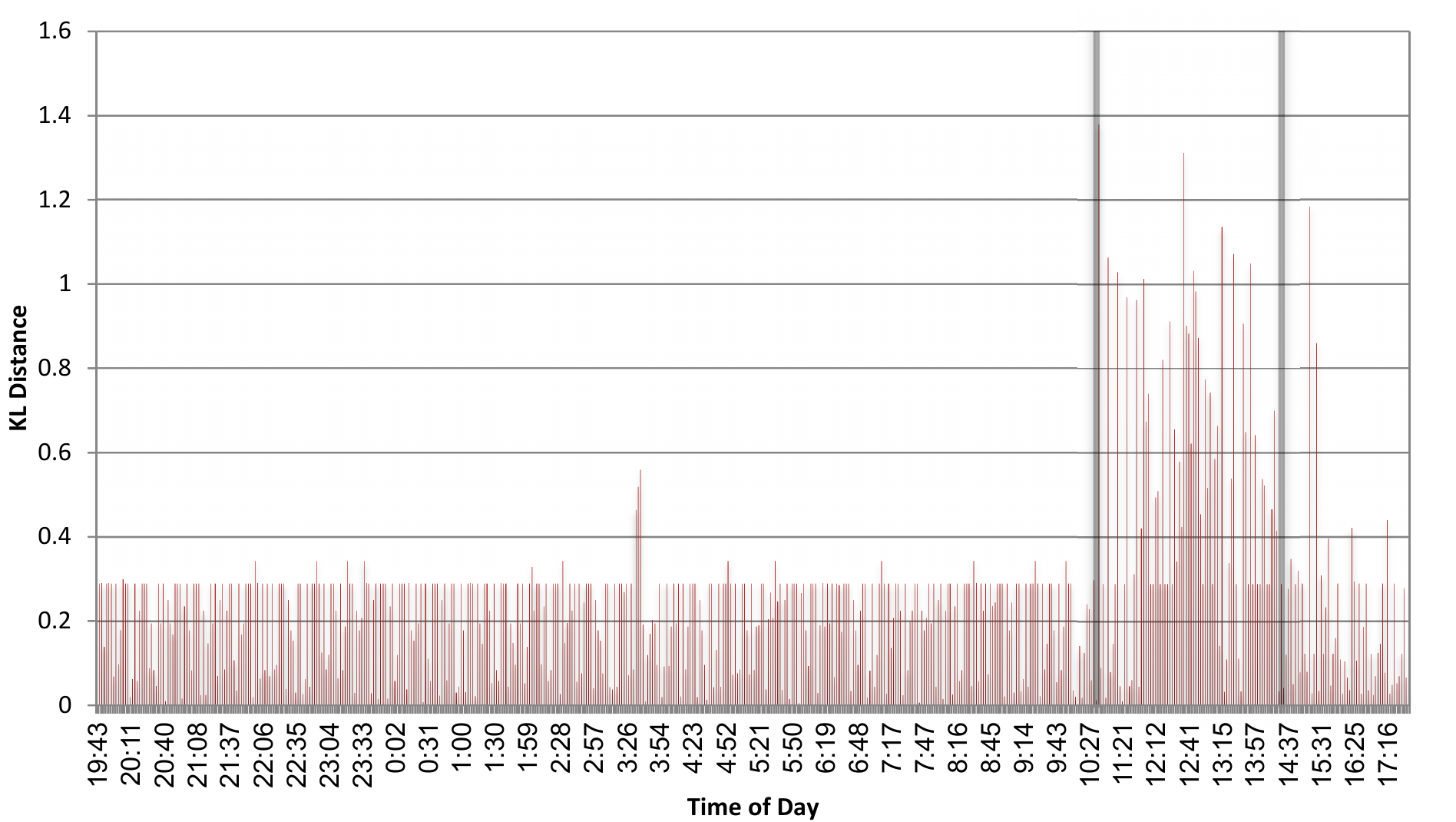}
\caption{KL distances between the learned histogram of normal behavior of firefox.exe and the histograms created every two minutes in the testing phase of the same process, as a function of time. The two gray vertical lines mark the time when ``Zeus" was active.}
\label{fig:firefox}
\end{figure*}

\Cref{fig:firefox} plots the KL distance between the histogram during the learning phase, and the histograms extracted during the testing phase. \emph{The process tested was firefox.exe, and the two vertical thick lines mark the time when the tool ``Zeus" was active}. It is very clear that the context of the system calls changes dramatically when the tool is active, and that simple monitoring of the KL distances every few minutes is sufficient to detect a change in the system behaviour.   
\subsection{Identifying Data Leakage}
In this part of the work, the universal anomaly detection algorithm was used in order to identify data leakage from a web server. Specifically, the setting was as follows. In the learning phase (a period of a few days), benign traffic on a web server was recorded using Wireshark \cite{wireshark}. Similar to the previous examples, timing-based sequences where extracted, quantized and used in order to build an LZ tree. This LZ tree served as a model for normal data.

Then, using Ncat \cite{ncat}, a script was installed on the server. This script initiated downloads of large chunks of data from the server. Several periods, each 30 minutes long, of traffic which includes Ncat were recorded. For comparison, similar length periods of traffic without Ncat were recorded as well. An LZ tree was built for \emph{each of the 30 minutes datasets}.

To identify data leakage, unlike the Botnets setting considered in \Cref{sec. res}, in this case, we compared the \emph{joint distributions of $k$-tuples} resulting from the LZ trees. That is, we used the distribution of $k$-tuples resulting from the LZ tree as an identifier for the data set, and calculated the distances between the distributions. 

\Cref{tab:leakage1} includes the results. The table depicts the distances between the learned, normal data, and $8$ testing periods, two which include data leakage using Ncat and $6$ without. Two distance measures where used: Mean Square Error (MSE) and KL distance. Under MSE, the leakage sessions clearly stand out compared to normal data. Results under the KL distance are less clear, especially in the first Ncast session, which included more normal data than the second.   
\begin{table}
\centering
\caption{Data leakage identification.}
\scalebox{0.72}{
	\begin{tabular}{ |l|l|l|l|l|l|l|l|l| }
	\hline
	& Ncat$_1$ & Ncat$_2$ & Normal$_1$ & Normal$_2$ & Normal$_3$ & Normal$_4$ & Normal$_5$ & Normal$_6$ 	\\ \hline
	MSE & 0.962 & 1.262 & 0.044 & 0.153 & 0.143 & 0.43 & 0.142 & 0.017 \\ \hline
	KL & 2.05 & 17.163 & 1.353 & 1.228 & 2.026 & 4.12 & 2.121 & 1.396	\\ \hline
	\end{tabular}	
}
\label{tab:leakage1}
\end{table}

Finally, to further challenge the algorithm, and see whether data leakage will also stand out when the normal communication includes (peaceful) massive downloads, the normal communication was augmented with benign downloads of various sizes. \Cref{tab:leakage2} depicts the results (under the KL distance). It is clear that while Ncat stands out compared to normal traffic on the web server, it is almost indistinguishable when the normal \emph{traffic learned includes downloads of large files}. This is expected, as Ncat uses a similar protocol, and the key differences in the timing are caused by file sizes. Hence, data leakage is clearly detected compared to normal surfing, yet, it is undistinguishable when the server, in peaceful times, serves large downloads.  
\begin{table}
\centering
\caption{Data leakage identification with additional downloads.}
\scalebox{0.82}{
	\begin{tabular}{ |l|l|l|l|l| }
	\hline
	& Normal & Normal + $1.3$MB & Normal + $10$MB & Normal + $200$MB \\ \hline
	Normal & 0.906&0.843&0.583&0.72\\ \hline
	\rowcolor[gray]{0.7} Ncat & 19.05 & 0.787 &0.733 &0.353	\\ \hline
	\end{tabular}	
}
\label{tab:leakage2}
\end{table}

\section{Discussion}\label{sec. discussion}

In this section, we discuss some related issues to the problem in question. First, we discuss what an attacker, the Botmaster, can do in order to neutralize our solution, and we show that it is quite complex to do so. Next, we present the infinite alphabet problem that exists when using the Lempel-Ziv compression algorithm as a probability assignment mechanism, and suggest another approach for that problem.

\subsection{What Can An Attacker Do?}

Remember that the statistical model constructed using the Lempel-Ziv algorithm is based on previously observed sequences, mainly generated by legitimate sources (clients), and each newly seen sequence is assigned a probability based on that model. Sequences with probability equal or higher than a predefined threshold are classified as normal, and otherwise classified as anomaly. 

Accordingly, the attacker may try to build the exact same model used by our suggested system in order to generate illegal sequences under the disguise of legitimate ones. For that matter, the attacker must have an access to the same database that was used to build the above statistical model. However, this sensitive information, in most cases, is protected and not available (that is, the service provider or organization implementing our system may use large amounts of legitimate traffic recorded at \underline{that} organization for the learning process).

Therefore, the attacker's strategy is to \emph{simulate that model}, or only part of it, where no prior knowledge of the underlying probability distribution of the sources generating the data sequences is available, by one of the following possibilities.

To be more precise and be able to quantify probabilities rigorously, we first define the following. Let the underlying alphabet be a binary alphabet $A=\{0,1\}$, and the length of the sequences be $n$. Assume legitimate sequences are generated $i.i.d$ according to an underlying probability $P$, which is unknown, and the attacker generates sequences $i.i.d$ according to a probability distribution $Q$, based on some estimate the attacker generated.    

In this case, the attacker may use a \emph{trial and error} strategy, and will randomly generate sequences according to probability distribution $Q$. Accordingly, the question arises is \emph{what is the probability to accept sequences that were generated according to Q?} To answer this question, we rely on the \emph{method of types} \cite{cover2012elements}, where a \emph{type} $P_{X}$ of a sequence $X = (x_{1}, x_{2}, \cdots, x_{n}$), $x_{i} \in A$, is defined as the relative proportions of occurrences of each element from $A$ in $X$ (which is a probability mass function of $X$ on $A$). For example, let $A=\{1,2,3\}$ and $X=12123$. Accordingly, the type $P_{X}$ is $P_X(1)=\frac{2}{5}, P_X(2)=\frac{2}{5}, P_X(3)=\frac{1}{5}$. A \emph{type class} of $P_{X}$, denoted as $T(P_{X})$, is the set of all sequences of length $n$ and type $P_{X}$ (for a more complete discussion, see \cite{cover2012elements}).

Under the above notation, the probability of type class $T(P)$ under distribution $Q^n$ is  $2^{-nD(P||Q)}$ to first order in the exponent, and more precisely, $\frac{1}{(n+1)^{|A|}}2^{-nD(P||Q)} \leq Q^{n}(T(P)) \leq 2^{-nD(P||Q)}$, where $D(P||Q)$ is the Kullback-Leibler Divergence measure (which acts as an error exponent component for that matter). Consequently, as long as the attacker does not know $P$, and uses an estimate $Q \neq P$, we have $D(P||Q) > 0$, hence the above probability $Q^{n}(T(P))$ decays \underline{exponentially} as $n$ grows to infinity. This means that, as we use longer sequences (in the testing phase), the attacker have less chance to bypass and neutralize our suggested solution with any estimate $Q \neq P$.   

Another attacking approach which needs to be considered is as follows. The attacker manages to obtain a legitimate sequence (that exists in the LZ78 phrase-tree) generated according to that $P$, e.g., by simulating/monitoring an HTTP legitimate connection and extracting the time differences from that session. First, the attacker may try to use it periodically by sending an attack sequence with the same pattern of the above sequence. This method will fail, as repeating a single sequence over and over again, even if it is legitimate and was derived from $P$, will create a stream whose distribution is far from $P$. For example, consider a case where one takes a short sequence, say $001$ of unbiased coin tosses, and generates $001001001\ldots$. Clearly, the resulting sequence will fail a test when comparing to an unbiased coin.

A more sophisticated approach is to generate new sequences based on the above available sequence as presented in \cite{merhav2004universal}. The basic idea is as follows. Given the above legitimate sequence, considered as a training sequence and denoted by $X^{m}$, where $m$ is the length of the sequence, and a string of $k$ purely random bits $U^{k}$, which are independent of $X^{m}$, the objective is to generate new sequence(s) of the same length or shorter ($n \leq m$), denoted as $Y^{n}$, with the same probability distribution of $X^{m}$ but with minimum statistical dependency between these sequences. That is, try to generate new sequences as if we had the generating source itself. To achieve this goal, a deterministic function $\phi(\cdot)$, independent of the unknown source $P$ is employed, such that $Y^{n}=\phi(X^{m},U^{k})$, and minimum mutual information $I(X^{m};Y^{n})$ is required in order to guarantee weak dependence, as much as possible, between the given training sequence and the result output sequences.  

However, from the results obtained, it follows that in order to faithfully represent the characteristics of the data, the input length $m$ must to be as large as possible, and the number of random bits $k$ needed to guarantee low dependency between $X^{m}$ and $Y^{n}$ grows linearly with the output length $n$.



Considering our problem domain, where the Botmaster generates new sequences according to the above model and updates its Bots, using the C\&C channels, with these sequences in order to carry out the attack. On the one hand, for the case where $n<m$, the Botmaster must constantly produce and maintain these $k$ random bits (to guarantee low statistical dependency), resulting in high complexity mechanism, and on the other hand, for the case where $n=m$, the Botmaster needs to generate large sequences (to preserve the characteristics of the original data, and specifically $P$), which may make it difficult to send these sequences, for example, as email attachments. Note, one may suggest that instead of receiving the above sequences through the C\&C channels, the Bots will generate them independently. This is disqualified due to both complexity (first to obtain a legal sequence and then to generate new ones, while Bots should operate in a simple manner as possible) and the requirement of a coordinated attack.


\section{Conclusions}\label{sec. conc}
In this work, we proposed a generic, universal anomaly detection framework. The proposed framework is based on universal compression and probability assignment, and it is able to build models for the learned data without any prior knowledge or model-assumptions. The models can then be used to detect anomalous behavior and alert in cases of attacks. 

Specifically, using universal probability assignment techniques based on the LZ-78 algorithm, we were able to suggest a modeling system which does not require any prior knowledge on the normal behavior, yet learns its statistical model optimally, in the sense that it converges to the true probability assignment whenever the source is stationary and ergodic. Together with the optimal decision rule, based on the Neyman-Pearson criteria, the probability assignments result in robust and efficient detection mechanisms. Moreover, as the technique suggested is based on practical universal compression, it can be implemented with low complexity and minimal pre-processing overhead.

To prove their applicability and test their performance, we applied the key techniques of this framework to several problems in computer security. The first was detecting C\&C channels of Botnets. We evaluated the system on \emph{real-world traces}. In particular, we offered to use \emph{time differences between events in the network} as the key feature, and showed how the \emph{context} of such a simple feature, easily learned using the suggested algorithm, enables the detection of most Botnets in the data set with a negligible false alarm probability. 

We continued with additional applications such as monitoring system calls in order to detect malicious tools and identifying data leakage. The results for these applications concurred with our main tests on C\&C detection, confirming the applicability of the suggested framework to several detection problems. Clearly, additional applications can be suggested. To name a few, we believe such tools can be used to identify abnormal behavior of users in computer systems or abnormalities in large data networks based on traffic patterns and communication partners of the tested nodes.
\bibliographystyle{IEEEtran}
\bibliography{bibliography}
\end{document}